# SPATIAL CORRELATIONS OF SI III DETECTIONS IN THE LOCAL INTERSTELLAR MEDIUM

M. Kuassivi

B.P. 170374 Cotonou, Bénin

ABSTRACT

Since the developpment of astronephography, we know that the Sun is about to exit the Local Interstellar Cloud (LIC). To date, because of its rare absorption signatures and the paucity of suitable neighbour targets, the LIC interface has proved to be elusive to extensive investigations. Comparing the spatial distribution of Si III detections found in the litterature along with 3 new sigtlines, I show that most detections seem to arise from a cone whose axis is parallel to the LIC heliocentric velocity vector. I interpret this result as an evidence that the heliosphere is actually interacting with the LIC frontier.

Keywords : Local Interstellar Cloud (LIC) - Conductive interfaces

## 1. Introduction

It has long been assumed that cooling in evaporative material at hot gas-cloud interfaces may dominate the radiative losses of a supernovae remnant expanding in a cloudy medium (McKee & Ostriker 1977) thus providing a global mechanism to explain enhanced EUV and X-ray backgrounds as well as UV absorption features of O VI. The generalization of this scenario to warm ISM clouds and cloudlets embedded in a hot medium has challenged the theorists for about three decades while extremely few observations are available to test the various models (Cowie & McKee 1977; Ballet et al. 1986; Boehringer & Hartquist 1987; Slavin 1989; Dalton & Balbus 1993; Slavin & Frisch 2002).

In a spectroscopic survey of 55 nearby white dwarfs using the IUE NEWSIPS SWP echelle data set, Holberg, Barstow & Sion (1998) report the detections of interstellar Si III lines in no less than 14 lines of sight. In a subsequent paper, Holberg et al. (1999), briefly discussed the presence of a highly ionized interface around cloudlets to account for the Si III ions; they were the first to suggest that different column densities are expected depending on the sightline geometry accross the interface.

In this paper, I analyse the spatial distribution of these Si III detections towards nearby stars, adding 3 new detections ($\varepsilon$CMA, $\alpha$Gru, and REJ 1032+532).

The observations and the data reduction are described in § 2, and a short discussion is given in § 3.

## 2. Data reduction

In the course of the present work, I use HST data obtained towards $\varepsilon$CMA, $\alpha$Gru, and REJ 1032+532. The STIS data used here have been collected through the MAST archive system. For data processing I followed the procedure described in Howk & Sembach (2000) combining IDL and IRAF treatments. For spectral regions covered in multiple orders, or when more than one observation is available, I co-added the flux-calibrated data weighting each spectrum by the inverse square of its error vector. The signal-to-noise ratio of the data is about 30 per pixel when the flux is at the level of the stellar continuum. Observations of the UV Si III obtained with the Ech-A and G160M gratings of GHRS are also used in this work towards $\varepsilon$CMA and $\alpha$Gru respectively. The details of the calibration can be found in Gry & Jenkins (2001) for $\varepsilon$CMA and in Redfield & Linsky (2000) for $\alpha$Gru.

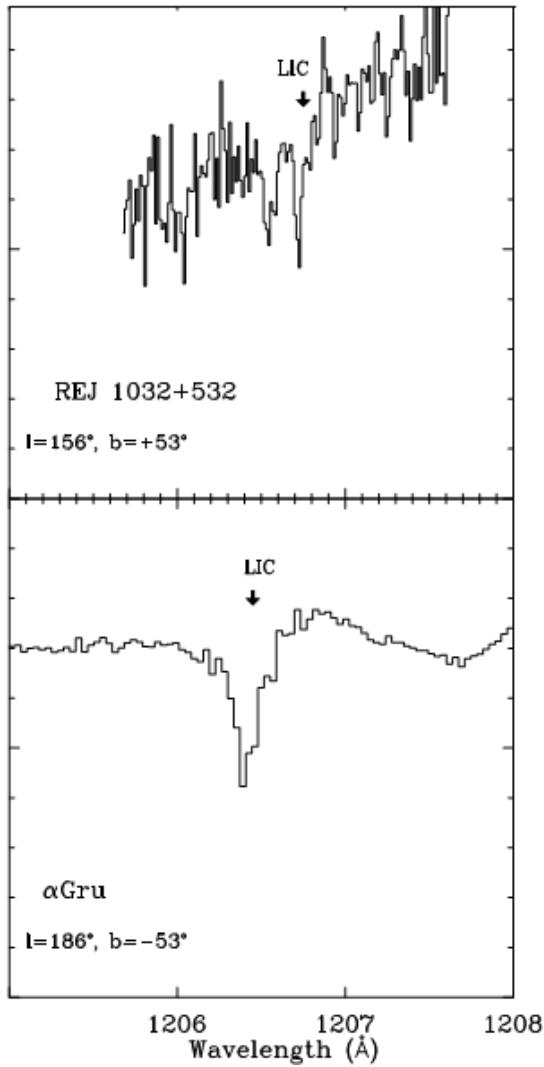

Fig 1. Detection of Si III towards REJ 1032+532 and αGru at the LIC velocity.

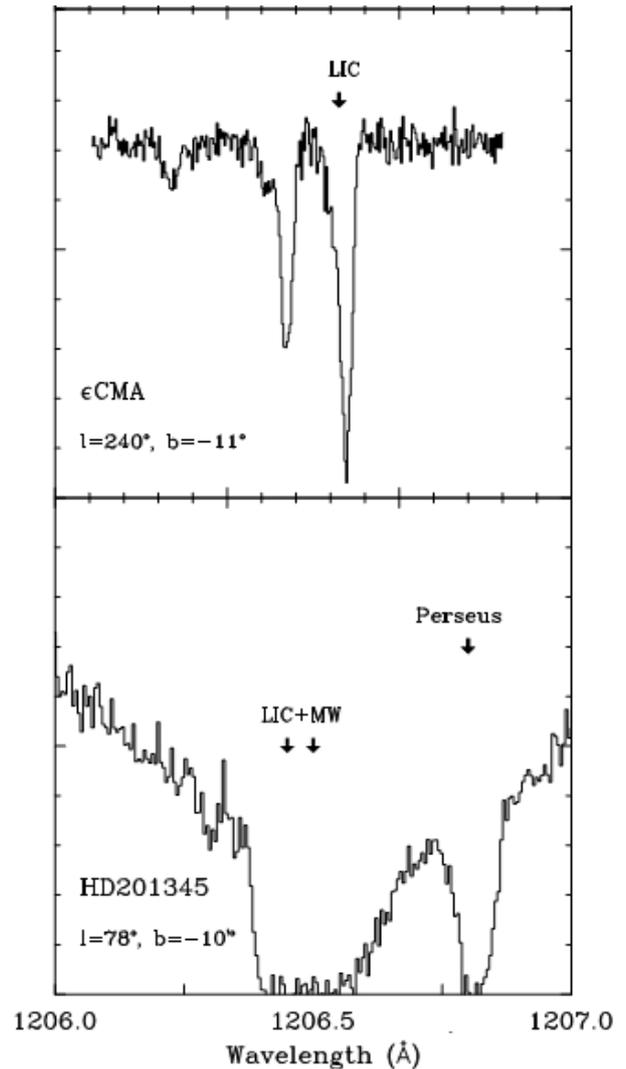

Fig2. Clear detection of Si III at the LIC velocity towards εCMA. Note that for longer sightlines (such as HD210345 at 1900 pc) the LIC component is usually blended.

## 3. Analysis and conclusion

Table 1. shows the Si III equivalent widths reported within the LIC. In order to avoid the blend with absorption lines from distant clouds, I discarded sightlines longer than 200 pc. Finally, using the detections reported by Holberg, Barstow & Sion (1998) and adding 3 new measurements towards αGru, εCMA and REJ, 1032+532 (see Fig.1 and Fig.2) we are left with a small sample of 11 detections.

Note that, in the particular case of the G191-B2B, I follow the analysis of Bannister et al. (2003) showing that most Si III detected arise within the G191-B2B strömgren sphere.

In order to point out a significant spatial correlation wihtin this data set, I used an iteration algorithm over the full sky area in search of detections clustering around a specific direction. If no spatial correlation were to be found, we expect data to be spread all over the x-axis of Fig.3 (from 0° to 180°).

As a matter of facts, Fig. 3 shows a clear correlation. With the exception of αGru, all the Si III detections arise within a narrow cone whose axis is defined by galactic longitude of 340° and galactic lattitude of -5°. This distribution mimicks a Si III trail with 30° opening angle following the Sun. Surprisingly enough, I note that the cone axis is about parallel to the LIC heliocentric downstream velocity vector (galactic longitude 351° and galactic lattitude +18°).

Because the Sun is on the edge of the LIC and that Si III is expected to arise within the LIC interface, we may consider the existence of a Si III trail as an evidence for Heliospheric – LIC magnetic interface interaction.

Indeed, the passage of the Sun trhough a magnetic interface may well have noticeable consequences in a near future. The influence of changes in solar Galactic environment on Earth is a long-standing question (Shapely, 1921). It is known that dramatic past cosmic events, such as the passage of a supernovae front shock or the crossing of a dense medium, do leave imprints on standard paleoclimatic records.

Recent paleoclimatic studies (Bard et al. 1997; Bond et al. 2001) have brought definitive evidences for a link between cosmogenic isotopes and the Earth's climate throughout the holocene.

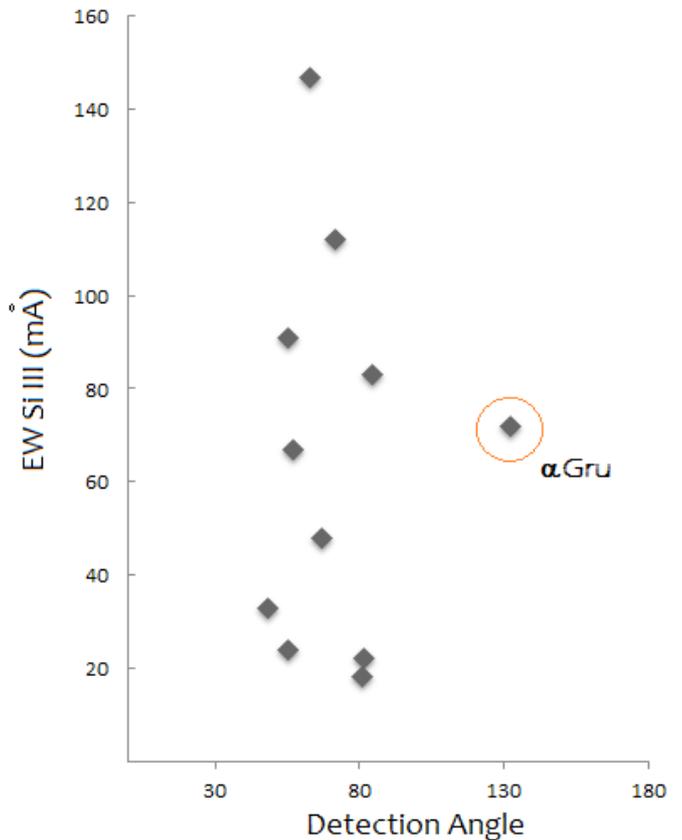

**Fig3.** Spatial correlation of Si III detections. Note that the inconsistent detection towards αGru is probably due to its sightline geometry crossing the LIC interface at low angle (Redfield and Linsky 2000).

| star | Distance (pc) | Si III EW(mÅ) | Angle relative to cone's axis |
|---|---|---|---|
| WD0004+330 | 97 | 67 | 57 |
| WD0232+035 | 74 | 24 | 56 |
| WD0455-282 | 102 | 112 | 71 |
| WD1202+608 | 200 | 91 | 55 |
| WD1234+481 | 129 | 48 | 67 |
| WD1800+685 | 159 | 147 | 63 |
| WD2309+105 | 79 | 22 | 82 |
| WD1034+001 | 100 | 83 | 85 |
| REJ 1032+532 | 132 | 33 | 48 |
| εCMA | 130 | 18 | 81 |
| αGru | 31 | 72 | 132 |
| G191-B2B (*) | 69 | 23 | 5 |

**Table 1.** Detections of Si III arising within 200 pc from the Sun. (*) Si III measured towards G191-B2B is circum-stellar (see Bannister et al. 2003).